\documentstyle[psfig]{mn}

\newcommand{\bd}{\begin{displaymath}}
\newcommand{\ed}{\end{displaymath}}
\newcommand{\be}{\begin{equation}}
\newcommand{\ee}{\end{equation}}

\title{Relation between radio core length and black hole mass for active
galactic nuclei}

\author[X. Cao and D. R. Jiang]
       {Xinwu Cao and D. R. Jiang \\
Shanghai Astronomical Observatory, National Astronomical Observatories,
Chinese Academy of Sciences,\\ Shanghai, 200030, China,
cxw@center.shao.ac.cn}
           
\date{Accepted . Received ; in original form}

\pagerange{\pageref{firstpage}--\pageref{lastpage}}
\pubyear{}

\begin{document}
\maketitle
\label{firstpage}

\begin{abstract}
We explore the relation between the linear length of radio core 
and the central black hole mass for a sample of radio-loud active
galactic nuclei (AGNs). An empirical relation between the size of
the broad line region (BLR) and optical luminosity is used to estimate
the size of the BLR. The black hole mass is derived from $\rm H_\beta$
line width and the radius of the BLR on the assumption that the
clouds in BLRs are orbiting with Keplerian velocities. A significant
intrinsic correlation is found between the linear length of the core
and the black hole mass, which implies that the jet formation is
closely related with the central black hole. We also find a strong
correlation between the black hole mass and the core luminosity. 
\end{abstract}

\begin{keywords}
galaxies:active-- galaxies:jets--quasars:general 
\end{keywords}

\section{Introduction}

Relativistic jets have been observed in many radio-loud AGNs and
are believed to be formed very close to the black holes. In currently
most favoured models of the formation of the jet, the power is generated
through accretion and then extracted from the disc/black hole rotational
energy and converted into the kinetic power of the jet (Blandford \&
Znajek 1977; Blandford \& Payne 1982). The disc-jet connection has been
investigated by many authors in different ways(Rawlings \& Saunders 1991;
Falcke \& Biermann 1995; Xu \& Livio 1999; Cao \& Jiang 1999; 2001).

The structure of jets in parsec scales in many AGNs has been revealed by
the Very Long Baseline Interfereometry (VLBI). 
The unresolved core might be the base of the jet, and
it is optically thick at the observed frequency (K\"onigl 1981), which can
successfully explain the fact that the radio emission from the core
usually exhibits a flat spectrum. For the nearby AGN M87, one can
even resolve their core-jet structure at a length scale of less than
100 Schwarschild radii at 43 GHz (Junor et al. 1999). VLBI observations
are useful to explore the physics at work in jet formation.

Recently, some different approaches are proposed to estimate the masses
of the black holes in AGNs, such as the gas kinematics near a hole (see
Ho \& Kormendy 2000 for a review and references therein). The
central black hole mass derived from the direct measurements on the gases
moving near the hole is reliable, but unfortunately, it is available
only for very few AGNs. For most AGNs, the velocities of the clouds 
in BLRs can be inferred from the widths of their broad lines.

If the radius of the BLR is available, the mass of the central black
hole can be derived from the broad-line width on the assumption that the
clouds in the BLR are gravitationally bound and orbiting with Keplerian
velocities (Dibai 1981), The radius of the BLR can be measured by using the
reverberation-mapping method from the time delay between the continuum
and line variations(Peterson 1993; Netzer \& Peterson 1997). Long-term
monitoring on the source is necessary for applying this method to derive
the radius of the BLR, which leads to a small amount of AGNs with measured
hole masses in this way. More recently, a tight correlation is found
between the size of the BLR and the optical continuum luminosity. One can
then estimate the size of the BLR in an AGN from its optical luminosity
(Wandel, Peterson \& Malkan 1999; Kaspi et al. 1996; 2000). The velocities of
the clouds in BLRs can be inferred from the broad line width. So, the
mass of the black hole in an AGN can finally be estimated from its
$\rm H_\beta$ width and the optical continuum luminosity(Laor 2000).

Many attentions have been paid on the black hole masses and accretion
types, and their relationship with radio emission (McLure \& Dunlop 1999;
Salucci et al. 1999; Laor 2000; Lacy et al. 2001; Gu, Cao \& Jiang 2001;
Ho \& Peng 2001; Ho 2001). Srianand \& Gopal-Krishna (1998) have explored
the correlation between the black hole mass and the largest linear radio
size for a sample of radio-loud quasars. They use the largest linear
radio size as a tracer of source age to study the evolution
the central black hole, which is different from our present work. 
In this work, we perform a statistical analysis
on a sample of radio-loud AGNs. The central black hole mass and the
linear length of the VLBI core are derived for all the sources in this
sample. In Sect. 2, we describe the sample. Sections 3-5 contain the
results and discussion. The cosmological parameters $H_0=75~
{\rm km~s^{-1}~Mpc^{-1}}$ and $q_0=0.5$ have been adopted in this paper.

\section{Sample}

Our starting sample is a combination of several large VLBI surveys.
It consists of: \\
1. Pearson-Readhead sample(Pearson \& Readhead 1988, hereafter
PR sample), The flux density limit at 5 GHz is 1.3 Jy;  \\
2. the first Caltech-Jodrell Bank VLBI survey (Xu et al. 1995,
hereafter the CJ1 sample). It extends the flux density limit of
PR sample from 1.3 to 0.7 Jy; \\
3. the second Caltech-Jodrell Bank VLBI survey (Taylor et al. 1994,
hereafter the CJ2 sample), Its flux density limit at 4850 MHz is 0.35 Jy.
A combination of flat-spectrum sources in PR sample, the CJ1 and CJ2
samples is given by Taylor et al. (1996); \\
4. a southern hemisphere VLBI survey of compact radio sources (Shen et al.
1997; 1998). Its flux density limit at 5 GHz is 1 Jy. \\
5. a large sample of radio reference frame sources observed by
the Very Long Baseline Array (VLBA) (Fey, Clegg \& Fomalont 1996; Fey \& Charlot 1997;
2000);\\
This combined sample contains more than 500 sources with both measured
core lengths and available redshifts. It has included most AGNs with
VLBI observations, though it is not a homogeneous sample. We believe that
it would not affect the main conclusions of this investigation.

The width of the broad line $\rm H_\beta$ is necessary in estimating
the central black hole mass. We focus on the sources in the starting
sample with redshifts in the range $0<z<1.5$.  This constraint limits
the starting sample to about 300 sources. We then search the literature,
collect all sources luminous than $M_{\rm B}=-23$ with available data
of $\rm H_\beta$ profiles.  Finally, we have a sample of 59
sources listed in Table 1.

\begin{figure}
\centerline{\psfig{figure=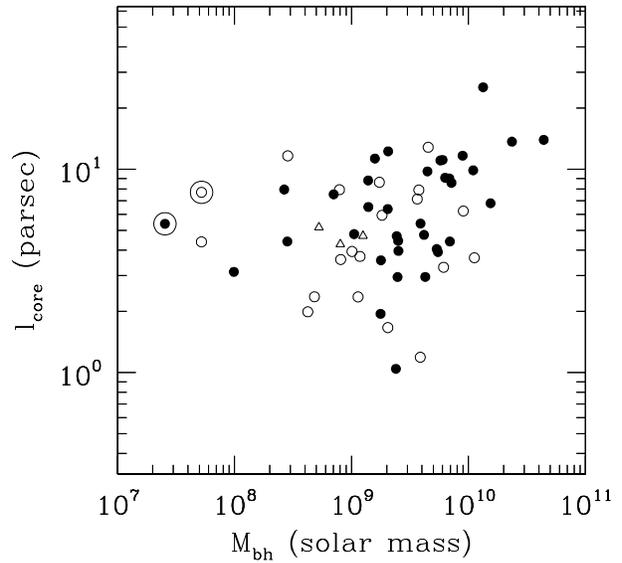,height=8.5 cm}}
\caption{The relation between the central black hole mass and the linear
length of the VLBI core corrected to the rest frame of the sources. The
full circles represent the sources observed at 8.55 GHz, and the open
circles represent the sources observed at 5 GHz, while the triangles
represent the sources observed at 2.32 GHz. The large open circles
represent the sources with FWHM($\rm H_\beta$)~$<~1000~{\rm km~s^{-1}}$. } 
\end{figure}

\begin{figure}
\centerline{\psfig{figure=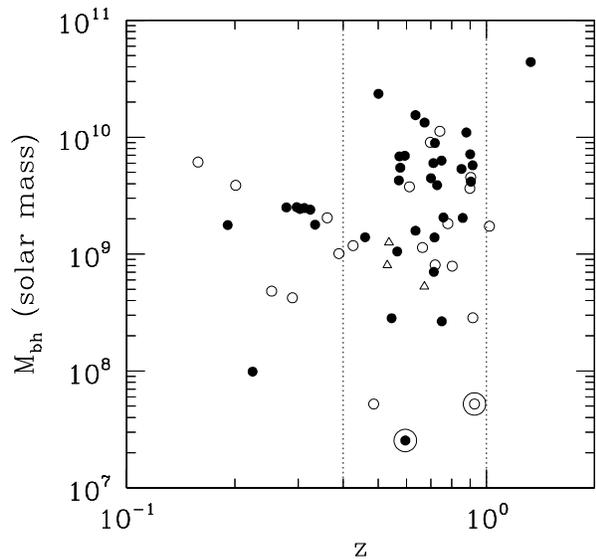,height=8.5 cm}}
\caption{The black hole mass versus redshift plane for the sample
(symbols as in Fig. 1).}
\end{figure}

\begin{figure}
\centerline{\psfig{figure=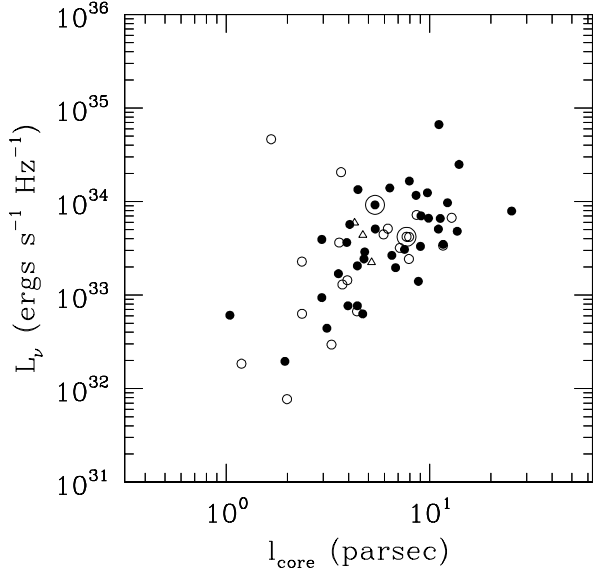,height=8.5 cm}}
\caption{The relation between the linear length of the VLBI core and
the core luminosity at 5 GHz (symbols as in Fig. 1). }
\end{figure}

\begin{figure}
\centerline{\psfig{figure=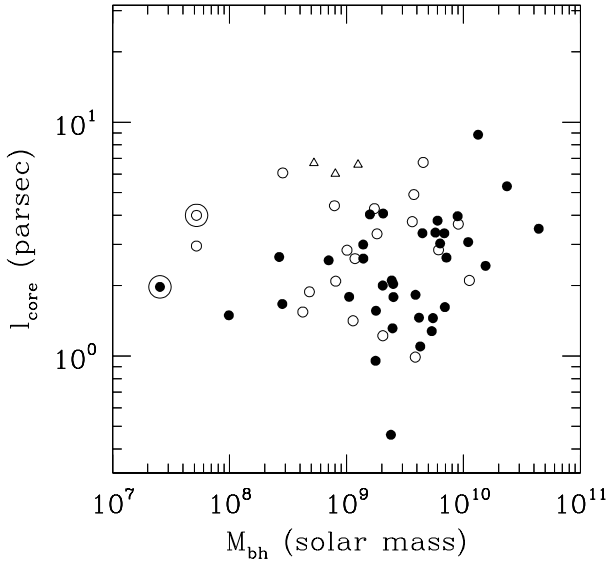,height=8.5 cm}}
\caption{Same as Fig. 1, but the linear length of the core has not
been correct to the rest frame of the sources.  }
\end{figure}

\begin{figure}
\centerline{\psfig{figure=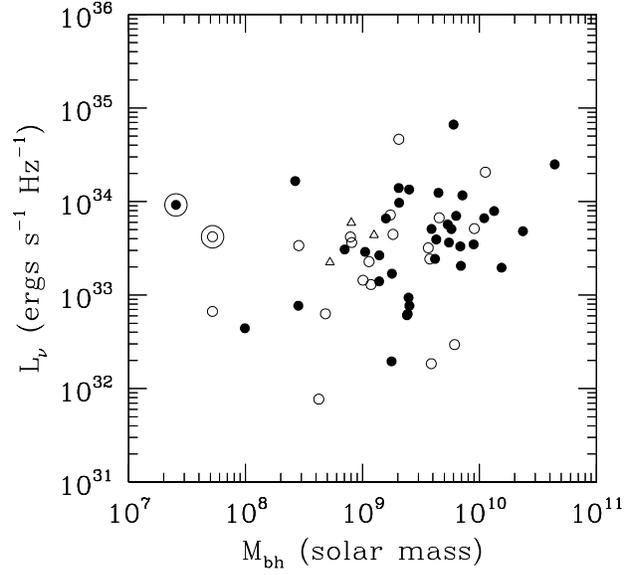,height=8.5 cm}}
\caption{The core luminosity at 5 GHz and the central black hole mass
relation (symbols as in Fig. 1). }
\end{figure}

\begin{figure}
\centerline{\psfig{figure=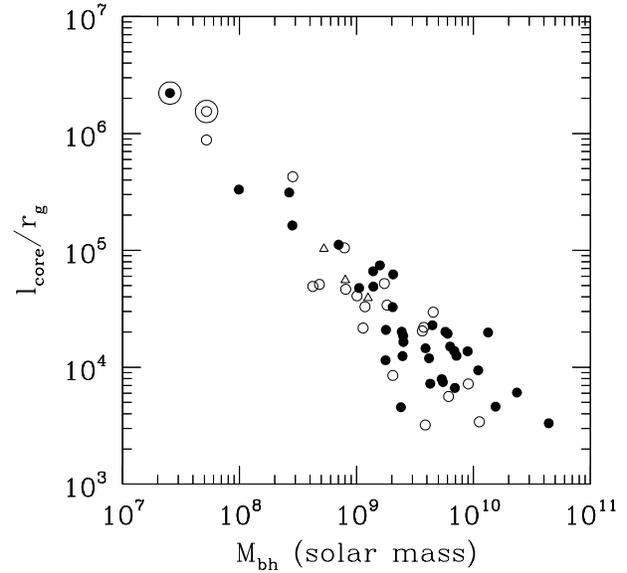,height=8.5 cm}}
\caption{Same as Fig. 1, but the length of the core is in unit of
the Schwarschild radius (symbols as in Fig. 1). }
\end{figure}

\section{Results}

The size of the BLR is derived by using the relation between
$R_{\rm BLR}$ and the optical continuum luminosity at 5100 \AA
(Kaspi et al. 2000). The velocity of the clouds in the BLR is
estimated from the full width at half maximum (FWHM) by:
$v=1.5 R_{\rm c}^{0.1}\times {\rm FWHM}({\rm H}_\beta)$
(McLure \& Dunlop 2001; Lacy et al. 2001), where $R_{\rm c}$ is the ratio
of the core to extended radio luminosity at 5 GHz in the rest frame of
the sources. The motion of the clouds may be anisotropic, and the
orientation effect would be important(McLure \& Dunlop 2001).
The parameter $R_{\rm c}$ is an indicator of the jet orientation.
The factor $R_{\rm c}^{0.1}$ is used to subtract the orientation effect
(Lacy et al. 2001). For those sources without measured $R_{\rm c}$,
we assume $R_{\rm c}=10$ and $R_{\rm c}=0.1$ for flat-spectrum and
steep-spectrum sources respectively (Lacy et al. 2001). The central
black hole masses can then be estimated on the assumption of the clouds
in BLRs orbiting with Keplerian velocities. All the data are listed
in Table 1.

The angular size of the core varying with the VLBI observing frequency
is common in compact extragalactic sources. It is found that the length
of the major axis of the core in the nucleus of M81 is proportional to
$\nu_{\rm obs}^{-0.8}$ (Bietenholz et al. 1996). For 3C345, multi-frequency
VLBI observations show that the angular size of the core is proportional
to $\nu_{\rm obs}^{-1.33}$ (Unwin et al. 1994). Recently, a statistic
analysis shows a similar power law relation: $\theta_{\rm core}\propto
\nu_{\rm obs}^{-1}$ for a large sample of AGNs (Jiang \& Cao, in
preparation). The dependence of the core size on the observing frequency
can be naturally explained in the frame of inhomogeneous sphere/jet models
(Marscher 1977; Blandford \& K\"onigl 1979; K\"onigl 1981). In this kind
of models, the radius at which optically thin emission can be seen from
the center varies with frequency as a power law. The sources in our present
sample are observed at three different frequencies, and redshifts of the
sources range from 0.158 to 1.327. So, we will correct the observed
length of the major axis of the core to the rest frame
of the sources. The length of the major axis $a_{\rm e}$ observed at
$\nu_{\rm e}$ in the rest frame of the sources is given by

\begin{equation}
a_{\rm e}=a_{\rm obs}\left({\frac {\nu_{\rm e}}
{\nu_{\rm obs}(1+z)}}\right)^{-\beta},
\end{equation}
where $\beta=1$ is adopted in this work (Jiang \& Cao, in preparation).

The linear length of the core is then available:

\begin{equation}
l_{\rm core}=a_{\rm e}D_{\rm a},
\end{equation}
where $D_{\rm a}$ is the angular diameter distance of the source.
We use this formula to convert the angular length to the linear length
observed at 5 GHz in the rest frame of the sources (listed in Table 1).

\subsection{Correlations between $l_{\rm core}$ and $M_{\rm bh}$}

The core lengths of the sources in our sample are observed at three
frequencies: 2.32 GHz, 5 GHz and 8.55 GHz, respectively. We use Eqs. (1)
and (2) to calculate the linear length observed at 5 GHz in the rest
frame of the sources.

We find that two sources in our sample have very narrow H$_\beta$ lines
(FWHM(H$_\beta$)$<$ 1000 km~s$^{-1}$). The widths of H$_\beta$ for these
two sources approximate to that of the typical narrow lines. It should be
cautious on the black hole mass estimate for these two sources, since
we cannot rule out the possibility that the H$_\beta$ line in these
two sources may be the narrow component emitted from the narrow line
regions (Gu et al. 2001). We therefore rule out these two
sources in all the following statistic analyses. 

The relation between the linear length $l_{\rm core}$
and the central black hole mass $M_{\rm bh}$ is plotted in Fig. 1.
A significant correlation is found between these two quantities
at 99.2 per cent confidence (Spearman rank correlation analysis).
The correlation coefficient is $r=0.351$.  We present the derived
black hole mass as functions of redshift $z$ for the sample in Fig. 2.
A weak correlation is found at 94 per cent confidence between the black hole
mass $M_{\rm bh}$ and redshift $z$. In order to explore whether the
correlation between $l_{\rm core}$ and $M_{\rm bh}$ is caused by the
common redshift dependence, we use the Spearman partial rank correlation
method (Macklin 1982) to check the correlation. The partial correlation
coefficient is 0.25 after subtracting the common redshift dependence.
The significance of the partial rank correlation is 1.86, which is
equivalent to the deviation from a unit-variance normal distribution
if there is no correlation present (Macklin 1982).
A summary of the results of partial rank correlation analyses is listed
in Table 2. We perform a correlation analysis on the sources in the
restricted redshift range $0.4< z <1$. For this subsample of 41 sources,
a correlation at 96 per cent confidence is still present between
$l_{\rm core}$ and $M_{\rm bh}$, while almost no correlation between
$M_{\rm bh}$ and $z$ is found (at 62 per cent confidence). It appears
that the correlation between $l_{\rm core}$ and $M_{\rm bh}$ is an
intrinsic one, not caused by the common redshift dependence.

{The relation between the linear length of VLBI core and
the core luminosity at 5 GHz (K-corrected to the rest frame of the
sources) is plotted in Fig. 3. A significant correlation is present
between these two quantities at 99.998 per cent confidence (the correlation
coefficient $r=0.575$). The common redshift dependence of these two
quantities has certainly played some roles in this correlation.
We use the Spearman partial correlation method to check the correlation.
The partial correlation coefficient becomes 0.254 after
subtracting the common redshift dependence (the significance of this
partial rank correlation is 1.892). The correlation coefficient is
0.51 independent of the black hole mass (the significance of the
correlation is 4.099). It implies that the correlation
between $l_{\rm core}$ and $L_{\nu\rm core}$ is an intrinsic one. 
The partial correlation coefficient of the relation between
$l_{\rm core}$ and $M_{\rm bh}$ is 0.177 independent of
core luminosity $L_{\nu\rm core}$. The significance of this
partial rank correlation is 1.302 (Macklin 1982). }

We then calculate the linear length of the core simply using
$l_{\rm core}^{\rm obs}=a_{\rm obs}D_{\rm a}$, i.e., without
any correction on the dependence of the core size on observing
frequency. The relation between $l_{\rm core}^{\rm obs}$ and $M_{\rm bh}$
is plotted in Fig. 4. For those 35 sources observed at the same
frequency 8.55 GHz, we find a correlation between $l_{\rm core}
^{\rm obs}$ at 97.1 per cent confidence.

\subsection{Correlation between $L_{\nu\rm core}$ and $M_{\rm bh}$}

We depict the relation between the VLBI core luminosity
(K-corrected to the rest frame of the sources)
and the central black hole mass in Fig. 5. A strong correlation is
present between $L_{\nu\rm core}$ and $M_{\rm bh}$
at 99.5 per cent confidence (the correlation coefficient $r=0.377$).
{The partial correlation coefficient is $0.287$ independent of
$l_{\rm core}$ (the significance of this partial correlation is $2.152$).
For the subsample of the sources in the restricted redshift range
$0.4< z <1$, a correlation is present between these two quantities
at 98 per cent confidence.}

\begin{table*}
 \begin{flushleft}
  \caption{Data of the core lengths and black hole masses.}
  \begin{tabular}{ccccccc}\hline
Source & z & $\nu_{\rm obs}$ & log($l_{\rm core}$) &
FWHM$ {\rm H_\beta}$ & Refs. & log($M_{\rm bh}/M_{\odot}$)\\
(1) & (2) & (3) & (4) & (5) &(6) & (7)\\ \hline
 0056$-$001 &  0.717 & 8.55 &  0.944 & 3000 &  B96  &  9.143 \\
 0110$+$495 &  0.389 & 5    &  0.595 & 3641 &  H97  &  9.004 \\
 0133$+$476 &  0.859 & 8.55 &  0.804 & 2697 &  L96  &  9.309 \\
 0227$+$403 &  1.019 & 5    &  0.935 & 1864 &  H97  &  9.239 \\
 0251$+$393 &  0.289 & 5    &  0.298 & 3152 &  VT95 &  8.626 \\
 0336$-$019 &  0.852 & 8.55 &  0.607 & 4876 &  JB91 &  9.728 \\
 0403$-$132 &  0.571 & 8.55 &  0.470 & 4780 &  C97  &  9.630 \\
 0405$-$123 &  0.573 & 8.55 &  0.955 & 3590 &  C97  &  9.836 \\
 0420$-$014 &  0.915 & 8.55 &  1.043 & 3000 &  B96  &  9.760 \\
 0444$+$634 &  0.781 & 5    &  0.773 & 3423 &  SK93 &  9.261 \\
 0518$+$165 &  0.759 & 8.55 &  1.088 & 4876 &  JB91 &  9.313 \\
 0538$+$498 &  0.545 & 8.55 &  0.644 & 1900 &  GW94 &  8.451 \\
 0554$+$580 &  0.904 & 5    &  1.108 & 3797 &  H97  &  9.656 \\
 0607$-$157 &  0.324 & 8.55 &  0.018 & 3518 &  H78  &  9.380 \\
 0724$+$571 &  0.426 & 5    &  0.571 & 2645 &  VT95 &  9.073 \\
 0730$+$504 &  0.720 & 5    &  0.555 & 2981 &  H97  &  8.908 \\
 0736$+$017 &  0.191 & 8.55 &  0.288 & 3400 &  B96  &  9.248 \\
 0738$+$313 &  0.635 & 8.55 &  0.832 & 4800 &  B96  & 10.189 \\
 0806$+$573 &  0.611 & 5    &  0.898 & 4135 &  H97  &  9.576 \\
 0830$+$425 &  0.253 & 5    &  0.372 & 2568 &  H97  &  8.683 \\
 0859$-$140 &  1.327 & 8.55 &  1.144 & 5700 &  N95  & 10.643 \\
 0923$+$392 &  0.698 & 5    &  0.794 & 7200 &  B96  &  9.956 \\
 0953$+$254 &  0.712 & 8.55 &  1.045 & 4012 &  JB91 &  9.778 \\
 0955$+$326 &  0.530 & 2.32 &  0.631 & 1380 &  B96  &  8.905 \\
 1012$+$232 &  0.565 & 8.55 &  0.680 & 2700 &  B96  &  9.022 \\
 1034$-$293 &  0.312 & 8.55 &  0.469 & 4101 &  S89  &  9.394 \\
 1045$-$188 &  0.595 & 8.55 &  0.731 &  622 &  S93  &  7.405 \\
 1058$+$629 &  0.664 & 5    &  0.372 & 2633 &  H97  &  9.056 \\
 1150$+$497 &  0.334 & 8.55 &  0.551 & 4810 &  B96  &  9.252 \\
 1151$+$408 &  0.916 & 5    &  1.065 & 1896 &  H97  &  8.455 \\
 1156$+$295 &  0.729 & 8.55 &  0.732 & 3700 &  B96  &  9.590 \\
 1226$+$023 &  0.158 & 5    &  0.517 & 3520 &  C97  &  9.787 \\
 1253$-$055 &  0.536 & 2.32 &  0.671 & 3100 &  WB86 &  9.099 \\
 1302$-$102 &  0.278 & 8.55 &  0.647 & 3400 &  B96  &  9.399 \\
 1309$+$555 &  0.926 & 5    &  0.886 &  704 &  H97  &  7.718 \\
 1354$+$195 &  0.719 & 8.55 &  1.066 & 4400 &  B96  &  9.950 \\
 1458$+$718 &  0.905 & 8.55 &  0.677 & 3000 &  B96  &  9.620 \\
 1510$-$089 &  0.361 & 5    &  0.221 & 2880 &  B96  &  9.310 \\
 1531$+$722 &  0.899 & 5    &  0.853 & 2764 &  VT95 &  9.562 \\
 1622$+$665 &  0.201 & 5    &  0.075 & 4579 &  VT95 &  9.588 \\
 1637$+$574 &  0.750 & 8.55 &  0.957 & 4620 &  N95  &  9.800 \\
 1641$+$399 &  0.593 & 8.55 &  0.644 & 3560 &  B96  &  9.841 \\
 1642$+$690 &  0.751 & 8.55 &  0.900 & 1845 &  L96  &  8.425 \\
 1656$+$053 &  0.879 & 8.55 &  0.994 & 5000 &  B96  & 10.040 \\
 1726$+$455 &  0.714 & 8.55 &  0.876 & 2953 &  H97  &  8.848 \\
 1734$+$363 &  0.803 & 5    &  0.899 & 3213 &  H97  &  8.897 \\
 1826$+$796 &  0.224 & 8.55 &  0.495 & 1059 &  H97  &  7.994 \\
 1856$+$737 &  0.460 & 8.55 &  0.814 & 4777 &  H97  &  9.144 \\
 1901$+$319 &  0.635 & 8.55 &  1.052 & 2580 &  GW94 &  9.200 \\
 1915$+$657 &  0.486 & 5    &  0.642 & 1204 &  H97  &  7.717 \\
 1928$+$738 &  0.303 & 8.55 &  0.671 & 3360 &  C97  &  9.386 \\
 2128$-$123 &  0.501 & 8.55 &  1.135 & 7050 &  C97  & 10.371 \\
 2143$-$156 &  0.701 & 8.55 &  0.989 & 4073 &  JB91 &  9.649 \\
 2155$-$152 &  0.672 & 2.32 &  0.714 & 1653 &  S89  &  8.721 \\
 2201$+$315 &  0.297 & 8.55 &  0.598 & 3380 &  C97  &  9.401 \\
 2216$-$038 &  0.901 & 8.55 &  0.933 & 3300 &  N95  &  9.854 \\
 2311$+$469 &  0.742 & 5    &  0.564 & 6385 &  SK93 & 10.050 \\
 2344$+$092 &  0.673 & 8.55 &  1.403 & 3900 &  B96  & 10.125 \\
 2345$-$167 &  0.576 & 8.55 &  0.592 & 4999 &  JB91 &  9.739 \\
                      \hline
\end{tabular}
\end{flushleft}
\end{table*}

\begin{table*}
\begin{minipage}{170mm}
Notes for the table 1. 
Column (1): IAU source name. Column (2): redshift.
Column (3): the observed frequency (in GHz);
Column (4): the core size corrected to the rest frame of the sources
at 5 GHz (in pc); 
Column (5): the FWHM of the $\rm H_\beta$ line (in km~s$^{-1}$);  
Column (6): references for the line widths; 
Column (7): the derived black hole masses \\

References:

B96: Brotherton (1996), 
C97: Corbin (1997), 
GW94: Gelderman \& Whittle (1994), 
H78: Hunstead et al. (1978), 
H97: Henstock et al. (1997)
JB91: Jackson \& Browne (1991), 
L96: Lawrence et al. (1996), 
N95: Netzer et al. (1995), 
S89: Stickel et al. (1989), 
S93: Stickel et al. (1993),  
SK93: Stickel \& K\"uhr (1993),
VT95: Vermeulen \& Taylor (1995),  
WB86: Wills \& Browne (1986). 
 \end{minipage}

\end{table*}

\begin{table*}
 \begin{minipage}{150mm}
  \caption{Spearman partial rank correlation analysis of the sample. 
Here $r_{\rm AB}$ is the rank correlation coefficient of the two variables,  
and $r_{\rm AB,C}$ the partial rank correlation coefficient. The
significance of the partial rank correlation is equivalent to the
deviation from a unit variance normal distribution if there is no
correlation present.
}
  \begin{tabular}{ccccc}\hline
Correlated variables: A,B & Variable: C & $r_{\rm AB}$ & $r_{\rm AB,C}$
& significance \\ \hline
$l_{\rm core}$, $M_{\rm bh}$ & z & 0.351 & 0.250 & 1.860\\
$l_{\rm core}$, $M_{\rm bh}$ & $L_{\nu{\rm core}}$ & 0.351 & 0.177 & 1.302\\
$L_{\nu{\rm core}}$, $M_{\rm bh}$ & z & 0.377 & 0.287 & 2.152\\
$L_{\nu{\rm core}}$, $M_{\rm bh}$ & $l_{\rm core}$ & 0.377 & 0.228 & 1.693 \\
$l_{\rm core}$, $L_{\nu{\rm core}}$ & z & 0.575 & 0.254 & 1.892\\
$l_{\rm core}$, $L_{\nu{\rm core}}$ & $M_{\rm bh}$ & 0.575 & 0.510 & 4.099\\
\hline
 \end{tabular}
\end{minipage}
\end{table*}

\section{Interpretation of the results}

K\"onigl (1981) proposed an inhomogeneous jet model, in which the magnetic
field $B(r)$ and the number density of the relativistic electrons ${n_e}(r,{%
\gamma _e})$ in the jet are assumed to vary with the distance from the apex
of the jet $r$ as $B(r)={B_1}(r/{r_1})^{-m}$ and ${n_e}(r,{\gamma _e})={n_1}%
(r/{r_1})^{-n}{\gamma _e}^{-(2\alpha +1)}$ respectively, and ${r_1}=1~{\rm pc}$.
If the bulk motion velocity of the jet is ${\beta }c$ (corresponding to a
Lorentz factor $\gamma $ ) with an opening half-angle $\phi $, and the axis
of the jet makes an angle $\theta $ with the direction of the observer,
the projection of the distance from the origin of the jet, $l_{\rm core}$,
at which the optical depth to the synchrotron self-absorption at the
observing frequency $\nu_{\rm e}$ in the rest
frame of the sources equals unity, is given by
equation (3) in K\"onigl (1981) as

\bd
l_{\rm core}=(2c_{2}(\alpha)r_{1}n_{1}\phi\csc
\theta)^{2/(2\alpha+5)k_m}(B_{1}\delta)^{(2\alpha+3)/(2\alpha+5)k_m}
\ed
\be
\times \sin \theta \nu_{\rm e}
^{-1/k_m}~~~{\rm pc},
\ee
where $c_{2}(\alpha)$ is the constant in the synchrotron absorption
coefficient, $\delta$ is the Doppler factor, and $k_{m}=[2n+m(2\alpha+3)-2]
/(2\alpha+5)$.

We assume that the mass loss rate of the jet is

\begin{equation}
\dot M_{\rm jet}=\eta_{\rm j}\dot M,
\end{equation}
where $\dot M$ is the accretion rate. We further assume that the magnetic
field pressure in the base of the jet can be scaled with the radiation
pressure of the disc at radius $r_{\rm d}$:

\be
{\frac {B^2(r_{\rm d})}{8\pi}}=
{\frac {8GM\dot M f}{3r_{\rm d}^3 c}} \eta_{\rm m}.
\ee
We assume $m=1$, and mass conservation in the jet, i.e., $n=2$.
Substituting Eqs. (4) and (5) into (3), we have

\bd
l_{\rm core}=\left [ 2C_2(\alpha)\alpha\eta_{\rm j}\pi^{-1}\phi
(1-\cos\phi)^{-1}\csc\theta \gamma_{e{\rm min}}^{2\alpha}
r_1^{-1}\beta^{-1}\right.
\ed
\bd
\left. \times c^{-1}\right]
^{2/{(2\alpha+5)k_m}}
\left[\left({\frac {3}{ {2\tilde r}_{\rm d}}}\right)^{1/2}
f^{1/2}\eta_{\rm m}^{1/2}r_{1}^{-1}c^{1/2}\right]^
{\frac {(2\alpha+3)}{(2\alpha+5)k_m}}
\ed
\bd
\times\delta^{\frac {2\alpha+3}{(2\alpha+5)k_m}}\nu_{\rm e}^{-1/k_m}
\left(1.11\times 10^{17}\eta^{-1}\dot m\right)^
{\frac {2\alpha+7}{2(2\alpha+5)k_m}}
\ed
\be
\times
\left({M\over {M_{\odot}}}\right)
^{\frac {2\alpha+7}{2(2\alpha+5)k_m}},
\ee
where $\dot m$ is the dimensionless accretion rate in unit
of Eddington rate, and ${\tilde r}_{\rm d}=r_{\rm d}/{\frac {2GM}{c^2}}$.
In this work, we have adopted $k_m=1$. It is
found that $l_{\rm core}\propto M^{0.64}$ for $\alpha=1$. 
We note that the index is insensitive to the value of $\alpha$.
The statistic results in Figs. 1 and 6 can therefore be interpreted
approximately by the inhomogeneous jet model. In fact, the linear
length of the core $l_{\rm core}$ is a function of several different
physical quantities, which may be the reason that the sources plotted
in Fig. 1 are dispersed over a large range.

\section{Discussion}

We find almost no correlation between the black hole mass $M_{\rm bh}$
and redshift $z$ in the restricted redshift range ($0.4<z<1$), while a
significant correlation is still present between the
linear length of the core $l_{\rm core}$ and the black hole
mass $M_{\rm bh}$ for this subsample. The possibility that the correlation
we found is caused by the common redshift dependence can be ruled out.
The Spearman partial correlation analysis confirms this conclusion. It is
therefore an intrinsic correlation.

We have to point out that the correlation between $l_{\rm
core}$ and $M_{\rm bh}$ might be affected by the angular
resolution limit of VLBI observations. The linear size of the core
is obviously a function of redshift, which is similar to the situation
for luminosity. If this is the case, this effect will be important
in the correlation analysis due to the common redshift dependence.
An analysis on the sources in a narrow restricted redshift range can
reduce this effect, since similar linear resolutions would be available
for all sources.  Our analyses show that the correlation between
$l_{\rm core}$ and $M_{\rm bh}$ is not caused by the common redshift
dependence. It seems that the effect of the angular resolution limit
of VLBI observations is not important in our analyses.

{It is not surprising that an intrinsic significant correlation
is present between the core length $l_{\rm core}$ and the core luminosity
$L_{\nu\rm core}$, since both the core length $l_{\rm core}$ and the core
luminosity $L_{\nu\rm core}$ scale with the magnetic field strength and
particle density in the jet (see K\"onigl 1981 and the discussion
in Sect. 4). These physical quantities of the jet might be closely
related with the central black hole. The black hole mass governs the
physical quantities of the jet, such as magnetic field and
particle density, and then the two observed quantities: the core
length $l_{\rm core}$ and the core luminosity $L_{\nu\rm core}$ (K\"onigl 1981).}

There is no significant correlation
between the linear length $l_{\rm core}^{\rm obs}$ (without frequency
correction to the rest frame of the sources) and $M_{\rm bh}$ for the
whole sample, while a correlation is present only for those sources
observed at the same frequency 8.55 GHz. The dependence of the core size on
observing frequency seems to be important in the correlation analyses,
which is in consistent with inhomogeneous sphere/jet models.

The linear core length $l_{\rm core}$ is a measurement on the size of
optically thick emission region of the plasma in inhomogeneous
sphere/jet models (Marscher 1977; K\"onigl 1981). The length $l_{\rm core}$
is mainly determined by the electron number density, the magnetic field
strength in the plasma and the bulk velocity of the plasma, 
which may be regulated by the central black hole mass. It is therefore
naturally to expect that a larger black hole has a larger radio emission
region. But the slope of the $M_{\rm bh}~-~l_{\rm core}$
is rather flat, i.e., the length $l_{\rm core}$ increases slowly
with $M_{\rm bh}$ (not a linear relation). We plot the relation
between the length  $l_{\rm core}$ in unit of the Schwarschild radius $r_{\rm g}
=2GM/c^2$ and the hole mass $M_{\rm bh}$ in Fig. 6. It shows that the
dimensionless linear length $l_{\rm core}/r_{\rm g}$ decreases with
the hole mass. As discussed in Sect. 4, the results seem to be consistent
with the inhomogeneous jet model. It is worth noting that a similar trend
of $l_{\rm core}/r_{\rm g}-M_{\rm bh}$ seems to be present for three Seyfert
galaxies observed by Ulvestad \& Ho (2001). It may imply that the origin
of the radio emission from these Seyfert galaxies is similar to that of
radio-loud AGNs. Their radio emission is generated by compact jets
(Falcke 1996; Falcke \& Markoff 2000), though the power of the jets in
these Seyfert galaxies is significantly lower than that in the radio-loud
AGNs discussed in present work.
 
It was pointed out that the core luminosity is a good indicator of jet
power for core dominated quasars (Cao \& Jiang 2001). The intrinsic
correlation found in this work between the core luminosity and 
hole mass confirms some previous similar results (McLure 1999; Laor 2000).
The jet formation in an AGN is regulated by the central black hole.

\section*{Acknowledgments}
We are grateful to the referee, Mark Lacy, for his helpful comments
and suggestions on the paper. XC thanks the support from  NSFC(No. 10173016), the NKBRSF
(No. G1999075403), and Pandeng Project. This research has made use of
the NASA/IPAC Extragalactic Database (NED), which is operated by the Jet
Propulsion Laboratory, California Institute of Technology, under contract
with the National Aeronautic and Space Administration. 

{}

\end{document}